\documentclass[a4paper,reqno,12pt,draft]{article}
\usepackage{amssymb,euscript}

\usepackage{epsfig}

\newcommand{\ie}{{\it i.e.}}
\newcommand{\be}{\begin{equation}}
\newcommand{\ee}{\end{equation}}
\newcommand{\bea}{\begin{eqnarray}}
\newcommand{\eea}{\end{eqnarray}}
\newcommand{\nn}{\nonumber\\}
\newcommand{\p}{\partial}
\newcommand{\ov}{\over}
\newcommand{\V}{{\rm Vol(X)}}

\newcommand{\ft}{\footnote}

\begin{document}

\begin{flushright}
\end{flushright}
\begin{flushright}
\end{flushright}
\begin{center}
\Large{\sc Flux and Freund-Rubin Superpotentials in M-theory}\\
\bigskip
{\sc Neil Lambert}\ft{lambert@mth.kcl.ac.uk}\\
{Dept. of Mathematics\\
King's College\\
The Strand\\
London\\
WC2R 2LS, UK\\}


\end{center}


\bigskip
\begin{center}
{\bf {\sc Abstract}}
\end{center}

We discuss the effective action
for weak $G_2$ compactifications of M-theory.
The presence of fluxes acts as a source for the 
the axions  and drives the Freund-Rubin parameter
to zero. The result is a stable non-supersymmetric vacuum with a 
negative cosmological constant.
We also give the superpotential which generates the effective potential
and discuss a simple model which aims to 
incorporate the effects of supersymmetry breaking by the gauge 
sector.

\newpage

\section{\sl Introduction}

Since the early days of  string theory it has been clear that
there is a critical need to  obtain phenomenologically realistic vacua. 
Originally this meant obtaining low energy effective actions which
agreed with the standard model or some supersymmetric/grand-unified 
generalization.
However over the past decade or so advances in cosmology 
have emphasized the point 
that string theory must also account for a cosmological 
standard model, incorporating  
inflation and a positive cosmological constant. 
Even more recently,  with the growing realization that string theory likely
contains a huge number of vacua \cite{Mike} - the so-called landscape - 
it has become important to gain an understanding
of the full structure of all four-dimensional 
vacua.  In other words we are not just interested in what we view as 
phenomenologically relevant but also what other scientists who live in 
other parts  of some great  multiverse would view as
phenomenologically relevant.

One approach to finding such vacua has been to consider 
M-theory  on singular special holonomy manifolds 
(for a recent review as well as a list of references see \cite{AG}).
The contribution of such M theory vacua to the landscape  has been recently
studied in \cite{ADV}. A subset of these constructions  revisits the 
Freund-Rubin ansatz within the context of singular weak $G_2$
manifolds. This program was detailed in  \cite{ADHL} and a few examples
were discussed. A benefit of this approach is that compact 
weak $G_2$ manifolds are easier to construct than compact $G_2$
manifolds. It is also possible to construct stable, chargeless 
brane configurations in the
compact space without the need to introduce orientifold planes.
On the other hand the four dimensional vacuum state is necessarily
anti-de Sitter space. 
In this paper we wish to study the low energy 
four-dimensional dynamics of 
these vacua, including internal  fluxes  from the 
M theory four-form.

Flux compactifications of M-theory to four dimensions, 
and in particular the role of the superpotential,  have been studied from 
a variety of points of view (for example see 
\cite{GVW,Sergei,AS,BW,Bobby,Minasian,Behrndt,Lukas,GL,HM}). 
In this paper we wish to consider the analysis of fluxes in compactifications
of $M$-theory on weak $G_2$ manifolds. We will mainly follow the analysis of
\cite{BW} for the case of $G_2$ compactification. Similar 
issues for weak $G_2$ compactifications have  
have recently been studied in \cite{HM}. 

A generic feature of 
anti-de Sitter space there is a mass splitting between scalar fields within
the same chiral supermultiplet. Thus one cannot simply truncate to the massless
modes and obtain a supersymmetric low energy effective action. Indeed it is
not clear in general that there is a supersymmetric truncation of an arbitrary
dimensional reduction involving a internal special holonomy  
Einstein manifold. We will use
the term moduli here to refer to any suitably light scalar field, which 
maybe massive or even tachyonic. 

A related problem that arises in ``compactifications'' over  an Einstein 
manifold is that the
mass scale of the light Kaluza-Klein modes is of the same order of magnitude as
the cosmological constant. Indeed Freund-Rubin solutions are perhaps more
naturally thought of as adS duals to  three-dimensional conformal 
field theories, rather than as traditional Kaluza-Klein models.
This presents a critical problem for phenomenological models built out of
a compactification on an Einstein space: one needs to somehow split the
mass scales so that the Kaluza Klein modes can be made sufficiently
heavy while  making the cosmological constant small. One of the 
motivations for the work presented here is to explore  
mechanisms, in particular supersymmetry breaking, where such a split 
might be made.

The rest of this paper is organized as follows. In section two we discuss
the four-dimensional low energy dynamics  of the ``light'' Bosonic modes. 
In particular we
derive an effective action for the metric and scalar fields.
In section three we postulate the form of the superpotential in
terms of geometrical data and show that it correctly reproduces the effective
potential of section two. In section four we consider a simple model which
attempts to incorporate the effect of supersymmetry breaking by gauge
theory fields which are localized at conical singularities in the internal
manifold. Finally in section five we close with a discussion and some
comments.

\section{\sl The Effective Potential}

Consider the Bosonic sector of the low energy effective action of
M-theory
\be
S = {1\over 2\kappa^9}\int \sqrt{-g_{11}} R - {1\ov 2}G\wedge\star G - {1\ov 6}C\wedge G\wedge G
\ee
where $G= dC$.
We are interested in compactifications of the form
\be
g_{11} = V_0\V^{-1} g_4({\bf M}) + g_7(X)
\label{gdef}
\ee
where
\be
\V = \int_X \sqrt{g_7}
\ee
is the volume modulus field
and $V_0$ is the volume of $X$ as measured in some solution that we
wish to perturb about. 

We will consider backgrounds that 
preserve four-dimensional $N=1$ supersymmetry and hence we
assume that $X$ is a weak $G_2$ manifold. This means that there exists
a spinor $\eta$ on $X$ such that
\be
\nabla_i\eta = {i\over 2}\lambda_7\gamma_i\eta
\label{Gstrucdef}
\ee
for some $\lambda_7$.
This in turn implies that
\be
R_{ij}(X) = 6\lambda_7^2g_{ij}(X)
\ee
From $\eta$ one can construct a three-form
\be
\Phi = {i\ov 3!}\bar\eta\gamma_{ijk}\eta dx^i\wedge dx^j\wedge dx^k
\ee
which satisfies
\be 
d_7\Phi = {4\lambda_7} \star_7 \Phi\ , \qquad  d_7\star_7 \Phi=0 
\label{threeformdef}
\ee
The existence of a three-form that satisfies (\ref{threeformdef}) provides an 
alternative definition of a weak $G_2$ manifold $X$.

To construct the four-dimensional effective action we follow \cite{BW}
and expand about a configuration which we take to be a  
Freund-Rubin flux background with
\be
G =\star_4 M + G_X\ , 
\label{Gback}
\ee
where  $\star_4$ is the four-dimensional  
Hodge star associated to $g_4$ and $G_X$ has no components tangent to ${\bf M}$. Here $M$ is the Freund-Rubin 
parameter, although it is important to keep in mind that it is not in general 
a constant, and $G_X$ is a topological flux, \ie\ a harmonic four-form.

Let us now discuss the light 
fields that will be present in the effective action. There will be
axion-like moduli $C^i$  from the periods of $C$ over three-cycles in $X$. 
As is usual for Kaluza-Klein theory the lightest such modes are in a one-to-one
correspondence with harmonic three-forms on $X$.

We will also obtain scalar moduli $s^I$ from the the moduli of $X$ which
preserve the existence of a weak $G_2$ form (\ref{threeformdef}), including
a possible rescaling of $\lambda_7$. These 
have been discussed in \cite{HM} and are in correspondence with three-forms
$\varphi^I$ that satisfy
\be
d_7 P_{\bf 1}\varphi_I  = 4\lambda_7\star_7 P_{\bf 1}\varphi_I\ ,
\qquad 
d_7 P_{\bf 27}\varphi_I  = -4\lambda_7\star_7 P_{\bf 27}\varphi_I\ ,
\qquad P_{\bf 7}\varphi_I=0
\label{smoduli}
\ee
where $P_{\bf n}$ denotes the projection onto the $n$-dimensional 
representation 
of $G_2$.

In the $G_2$ holonomy case where $\lambda_7=0$ both these moduli are in a
one-to-one correspondence with  harmonic three-forms and together form
the complex Bosonic scalar of a four-dimensional chiral supermultiplet.
If $\lambda_7\ne 0$ then this is not the case. 
Indeed in the classic Freund-Rubin example where
$X= S^7$ there are no harmonic three-forms and
yet there is a volume modulus which preserves the weak $G_2$ structure,  
up to a rescaling of $\lambda_7$.

Thus it follows that we should consider two types of complex moduli 
$z^i = C^i+i\tilde s^i$ and $z^I = \tilde C^I + is^I$. 
The $C^i$ and $s^I$ are massless (at least when the Chern-Simons and Flux 
terms are ignored)  whereas
their superpartners $\tilde s^i$ and $\tilde C^I$ will not be. 
This mass splitting between scalars within a supermultiplet is what
one expects for chiral  supermultiplets in anti-de Sitter space.

The $\tilde C^I$ have a straightforward 
interpretation as massive Kaluza-Klein modes.  
In particular our full ansatz for the three-form $C$ is
\be
C = \sum_i C^i\omega_i  +\sum_I \tilde C^I\varphi_I+ C_{X}
+ C_0
\label{KKC}
\ee
here $\omega_{i}$ are basis of harmonic three-forms on $X$ and 
$C_{X}+C_0$ is the background $C$-field that gives rise to (\ref{Gback}).
The $C^i$ and $\tilde C^I$ are therefore massless and massive 
Kaluza-Klein modes respectively.
Note that we have assumed that there are no one-cycles on $X$
and we set the vector fields $A$, which arise from harmonic two-forms 
on $X$,   to zero.

The precise role of
the $\tilde s^i$ moduli is less clear. 
 They  correspond to
deformations of $\Phi$ by  harmonic three forms
and hence they do not preserve the  weak $G_2$ structure. 
Thus we are led  to parameterize the weak $G_2$
form by
\be
\Phi = \sum_I s^I\varphi_I +\sum_i\tilde s^i\omega_i
\label{KKPhi}
\ee
It should be kept in mind that we expand about a configuration
with $s^I\ne 0$ and $\tilde s^i=0$ 
and under  
a generic deformation involving the $\tilde s^i$ the internal manifold 
$X$ is no longer a weak $G_2$ manifold. In this section we will simplify
our calculations by  setting $\tilde s^i=0$. We can readily
deduce the kinetic terms for the $\tilde s^i$ since they are related by 
supersymmetry to the kinetic terms for  $C^i$. 
Since the $\tilde C^I$ are more massive than their superpartners,
and the $C^i$ are massless, we expect that the $\tilde s^i$ are
tachyonic.
In the next section we will  
deduce the potential for $\tilde s^i$ from supersymmetry and verify that
this is the case.

Note that in the limit $\lambda_7\to 0$ we recover two copies
of the supermultiplets found in a $G_2$ compactification. 
In other words turning on $\lambda_7$ introduces a splitting between
two identical copies of the same multiplet. This seems odd as one might have
expected a smooth limit as $\lambda_7\to 0$. The cause of this pathology
can be traced to the fact that if $\lambda_7\ne 0$ then the
$\omega_i$ and $\varphi_I$ forms are orthogonal
\be
\int_X \omega_i\wedge \star \varphi_I
\sim {1\over \lambda_7}\int_X \omega_i\wedge d\varphi_I =0 
\ee
but this is not the case if $\lambda_7=0$. Hence one can't really think of
weak $G_2$ manifolds that are close to being $G_2$ by taking $\lambda_7$ small. This shouldn't be
surprising as by blowing up the volume we can send $\lambda_7 \to 0$ but
this clearly does not change the manifold in any interesting way.
  
Our first step is to  
substitute this ansatz into the original action and obtain the following 
effective action in the Einstein frame:
\be
S_{eff} = {1\over \kappa^2_4}\int \sqrt{-g_{4}} \left(\frac{1}{2}R_4 
-g_{i\bar  j}\p_\mu z^i\p^\mu \bar z^{\bar j} 
-g_{I\bar  J}\p_\mu z^I\p^\mu \bar z^{\bar J}
-   V_{eff}\right) + {\cal T} +\ldots 
\label{wrongEA}
\ee
Here the ellipsis denotes terms involving vector fields and Fermions and
\be
\kappa^2_4 = {\kappa^9\ov V_0}
\ee 
is the four-dimensional Planck length.
One can  see that there is no cross kinetic term $g_{i\bar J}$ as this
would imply a term $\p_\mu C^i\p^\mu \tilde C^J$ which vanishes
since Kaluza-Klein modes of different levels  are orthogonal.
In particular we find
\bea
g_{i\bar j} &=& {1\over 4\V}\int_X \omega_i\wedge \star_7 \omega_j\nn
g_{I\bar J} &=& {1\over 4\V}\int_X \varphi_I\wedge \star_7 \varphi_J\nn
V_{eff} &=& 16 \lambda_7^2\frac{V_0}{\V}\tilde C^I\tilde C^Jg_{IJ}
-{21  V_0\lambda_7^2\ov \V} +\frac{\V^3}{4\cdot 4! V_0^3}G_0\wedge\star_4G_0\nn
&&
 + \frac{1}{4}{V_0\ov \V^2}\int_X G_X\wedge \star_7 G_X \nn
{\cal T}&=&-{1  \ov  4 V_0} G_0 C^i\int_X \omega_i\wedge G_X
-\frac{1}{4V_0} G_0 \tilde C^I\tilde C^J\int \varphi_I\wedge d\varphi_J\nn
&&
- {1  \ov  4 V_0} G_0\int_X C_X\wedge G_X\nn
\eea
Note that we
have written $G_0$  instead of $\star_4M$ to emphasize that the Chern-Simons
term is independent of the metric as well 
as to make the the metric dependence of
the $G_0\wedge\star_4G_0$ term more explicit.
Note also that there is no linear term in $\cal T$ from the 
Kaluza-Klein axions $\tilde C^I$ since,
if $\lambda_7\ne 0$, 
\be
\int_X \varphi_I\wedge G_X = \int_X \star_7\varphi\wedge \star_7 G_X
 \sim\frac{1}{\lambda_7}\int d \varphi\wedge \star_7 G_X =0 
\ee
as we have taken $G_X$ to be harmonic.

First consider the case without fluxes but $\lambda_7\ne 0$. 
If we vary the volume, say for example by rescaling $g_7\to \Omega^2 g_7$,
then it is clear that $\lambda_7\to \Omega^{-1}\lambda^7$ and hence
\be
\V {\p \over \p \V}\lambda_7 = -\frac{1}{7}\lambda_7
\ee
Similarly we note for future reference that
\be
\V {\p \over \p \V}\int_X G_X\wedge \star G_X  = -{1\over 7}\int_X G_X\wedge \star_7 G_X
\ee
We find an 
extremum of $V_{eff}$ at
\be
\V = \left({36\lambda_7^2\over M^2}\right)^{1\ov 4}V_0\ , 
\qquad V_{eff} = 
-5\sqrt{6}\lambda_7^\frac{3}{2}M^\frac{1}{2}
\ee
This  is not the Freund-Rubin solution. While
this potential predicts the correct value for $\V$ it does not
reproduce the correct value for the cosmological constant. Thus we see
that the effective action (\ref{wrongEA}) is misleading, even at
the classical level. We will see that
the cause is that $M$ is not a constant. However it can be integrated out
using the equations of motion, in the form of the conservation of Page charge.

To see this in more detail we can proceed with our  analysis 
in the presence of fluxes. We see 
from the effective action  that there will be a 
tadpole for the massless axion fields coming from
the  Chern-Simons term in the potential
\be
{\p {\cal T} \ov \p C^i} = -{1  \ov 4 V_0} G_0\int_X \omega_i \wedge G_X
\ee
Indeed this prediction can be verified 
directly using the full eleven dimensional equation of motion 
$d\star G+\frac{1}{2}G\wedge G=0$.
If we assume that $G = G_0 + G_X$, \ie\ the axions are constant, 
then projecting this equation onto the components tangent to
${\bf M}\times \Sigma$, where $\Sigma$ is a four-cycle in $X$
gives
\be
d\star G_X + MG_X =0
\ee
Thus $G_X$ is exact and hence there can be no topologically non-trivial  
fluxes if $M\ne 0$.\footnote{I am grateful to B. Acharya and
F. Denef for pointing this out.} This
forbids static solutions with both topologically non-trivial 
fluxes and a non-vanishing 
Freund-Rubin parameter $M$. 
What we have seen is that turning on both a Freund-Rubin parameter
and flux acts as a source for the axions.

A related observation is that $M$ is no longer conserved in the
presence of fluxes and non-trivial axions. Rather the conserved the
object is the Page charge
\bea
P_0 &=& \int_X \star G + {1\over 2}C\wedge G \nn 
&=&  {\V^3\over V^2_0}\star_4 G_0 
+{1\over 2} C^i \int_X \omega_i\wedge G_X 
+ {1\over 2} \int_X C_X\wedge G_X\nn&&
+\frac{1}{2}\tilde C^I\tilde C^J\int_X \varphi_I\wedge d\varphi_J\nn
\label{Page}
\eea
Hence $M$ alone is not conserved. 
Therefore we should 
take into account the full  $C_0$ equation of motion which will result in 
a time-dependent $M = -\star_4G_0$. To proceed we write
\be
\star_4G_0 = \tilde P_0{V_0^2\over \V^3} 
- {1\over 2}{V_0^2\over \V^3} C^i \int_X \omega_i\wedge G_X 
-\frac{1}{2}\frac{V_0^2}{\V^3}  \tilde C^I\tilde C^J \int_X \varphi_I\wedge d\varphi_J
\label{Mis}
\ee
where we have redefined the Page charge  to absorb a constant
arising from the fluxes 
\be
\tilde P_0 = P_0 - {1\over 2}\int_X C_X\wedge G_X
\ee

We can now substitute (\ref{Mis}) into the remaining
equations of motion. This leads to a system of equations for the
metric and moduli which  arise  from the action
\be
S_{eff} = {1\over \kappa^2_4}\int \sqrt{-g_{4}} \left(\frac{1}{2}R_4 -g_{i\bar
  j}\p_\mu z^i\p^\mu \bar z^{\bar j} -  
g_{I\bar  J}\p_\mu z^I\p^\mu \bar z^{\bar J} -U_{eff}\right)  
\label{actionfinal}
\ee
with 
\bea
U_{eff} &=& \frac{16 \lambda_7^2V_0}{\V}\tilde C^I\tilde C^Jg_{IJ}  -{21  V_0\lambda_7^2\ov \V}
 + \frac{1}{4}{V_0\ov \V^2}\int_X G_X\wedge \star_7 G_X \nn
&&
+{V_0\over 4 \V^3}\left(\frac{1}{2}C^k \int_X \omega_k\wedge G_X +
\frac{1}{2}\tilde C^I\tilde C^J \int_X \varphi_I\wedge d\varphi_J
- \tilde P_0\right)^2\nn
\eea
Note that the final term within the brackets is proportional to $M$.
Note also that this action does not simply result from substituting (\ref{Mis})
into the effective action (\ref{wrongEA}), except if $P_0=0$. 
Thus we find that classically we can ``integrate out'' the Freund-Rubin
parameter using its  equation of motion, \ie\  conservation of
Page charge, 
and find an effective action for the remaining light fields
involving $U_{eff}$. 
The effective potential $U_{eff}$ is a generalization of that
found in  \cite{BW} to include $\lambda_7$ and
a non-vanishing $P_0$. It also generalizes the result of \cite{HM} to
include fluxes and the additional $z^i$ moduli.

We now  recover the correct Freund-Rubin solution if the fluxes vanish.
It occurs at
\be
\V = \left({36\lambda_7^2\over M^2}\right)^{1\ov 4}V_0\ , 
\qquad U_{eff} = -2\sqrt{6}\lambda_7^\frac{3}{2}M^\frac{1}{2}
\label{FR}
\ee
and contrary to the maximum of $V_{eff}$ found above this does
have the correct value of the cosmological constant.
Thus the correct 
effective action is the one containing $U_{eff}$, \ie\ with 
$M$ integrated out in favour of $P_0$. This shows that, even at the
classical level, the effective potential $V_{eff}$ is incorrect
and one needs to remove $M$ by its equation of motion to
obtain a valid effective potential for the scalar moduli alone.

The effective potential $U_{eff}$ is bounded from below and has a
global minimum corresponding to $\tilde C^I=0$ and  
$M=0$ with negative energy.  Since the minimum is only one
constraint we see that there will also be $b_3-1$ flat directions
corresponding to axionic modes which are perpendicular to 
\be
\int_X \omega_i\wedge G_X 
\ee
Therefore if one starts out with a Freund-Rubin parameter $M$ and topologically
non-trivial fluxes then the system
will evolve until $M$ is driven to zero.

Finally we would like to comment that the Page charge is somewhat
mysterious as, on the one hand it is quantized, and yet under
a large gauge transformation $C_X \to C_X + \Omega$ we see that
\be
P_0 \to P_0 + {1\over 2}\int_X \Omega\wedge G_X
\ee
Thus $P_0$ is only gauge invariant modulo an integer and in some
cases can be set to zero
(for comments on this see \cite{BW} and \cite{Greg}). 
However this is not possible in the absence of
fluxes, since in this case $P_0$ is 
gauge invariant. Indeed 
in the absence of fluxes we should be able to identify
the Freund-Rubin solution as an extremum of $U_{eff}$ and 
we have seen that this requires 
$P_0\ne 0$ in order to stabilize $\V$.

\section{\sl The Superpotential}

We wish to cast the above potential in the generic form for $N=1$
supergravity
in terms of the Kahler potential $K$ and a superpotential $W$
\be
U_{eff} = e^K\left(g^{i\bar j}D_iW D_{\bar j}\bar W+g^{I\bar J}D_IW D_{\bar J}\bar W - 3W\bar W\right)
\ee
with $D_i W = \p_iW + \p_iKW$ and $\p_i = \partial/\partial z^i$ and
similarly for $i \to I$. This will also allow us to deduce the dependence on the
metric moduli $\tilde s^i$.
For the case that $\lambda_7=0$ the superpotential was derived in
\cite{BW} and for the case $G_X=0$ it was derived in \cite{HM}.
Here we wish extend these results to our case with both $\lambda_7$
and $G_X$ non-vanishing.

The general form for  superpotentials in the presence of fluxes 
was first proposed in \cite{GVW,Sergei}. For the $G_2$ case 
the superpotential was taken in \cite{BW} to be
\be
W = \frac{1}{4V_0}\int_X \left(\frac{1}{2}C+i\Phi\right)\wedge G
\ee
however one can easily check that this is no longer holomorphic 
if $d\Phi \ne 0$.
A natural choice for the generalization is (see also \cite{AS,HM,Ruben}) 
\be
W =-\frac{1}{4V_0}P_0+ \frac{1}{8V_0}\int_X \left(C+i\Phi\right)\wedge d\left(C +i\Phi\right)
\ee
and we will show that this does indeed reproduce the effective potential
$U_{eff}$, along with a suitable choice of Kahler potential.
From this expression it is clear that $W$ is holomorphic
\be
\delta W 
=\frac{1}{4V_0}\int_X \left(\delta C+i\delta \Phi\right)\wedge d\left(C +i\Phi\right)\nn
\ee
provided that we view $P_0$ as a constant, rather than given by the
expression (\ref{Page}). This is reasonable as our effective action is
only valid within a supersection sector where the Page charge is held fixed.
Furthermore we see that 
\bea
W 
&=& -\frac{1}{4V_0}P_0+\frac{1}{8V_0}\int_X \left(C+i\Phi\right)\wedge G  +
\frac{i}{8V_0}\int_X \left(C+i\Phi\right)\wedge d\Phi\nn
&=&-\frac{1}{4V_0}P_0+\frac{1}{4V_0}\int_X \left(\frac{1}{2}C+i\Phi\right)\wedge G- \frac{1}{8V_0}\int_X \Phi\wedge d\Phi\nn
\eea
so if $d\Phi=0$ then we recover the original superpotential of \cite{BW}, 
shifted by the addition of the constant $P_0$.

Our first step is to calculate $\p_i W$ and $\p_I W$. Here we find
\be
\p_i W = \frac{1}{4V_0}\int_X \omega_i\wedge G_X \qquad 
\p_I W = \frac{1}{4V_0}\tilde C^J\int_X\varphi_I\wedge d\varphi_J +\frac{i}{4V_0}\int_X \varphi_I\wedge d\Phi
\ee
Recall that we are using the Kaluza-Klein ansatz (\ref{KKC}) as
well as an expansion of $\Phi$ as in (\ref{KKPhi}). Following similar arguments
to those in \cite{BW} we can then see that
\bea
g^{i\bar j}\p_i W\p_j\bar W &=&\frac{\V}{4V_0^2} \int_X G_X\wedge \star G_X\nn
g^{I\bar J}\p_I W\p_J\bar W &=&\frac{16\V^2\lambda_7^2}{V_0^2}\tilde C^I\tilde C^Jg_{IJ}
+\frac{1}{16V_0^2}g^{I\bar J}\int_X \Phi_I \wedge d\Phi \int_X \Phi_J \wedge d\Phi\nn
W&=&  -\frac{1}{4V_0}
\tilde  P_0 + \frac{1}{8V_0}C^i\int_X \omega_i\wedge G_X  
+\frac{1}{8V_0}\tilde C^I\tilde C^J\int_X \varphi_I\wedge d\varphi_J
\nn&& -\frac{1}{8V_0}\int_X \Phi\wedge d\Phi
 + \frac{i}{4V_0}\tilde C^I\int_X
\varphi_I\wedge d\Phi
+\frac{i}{4V_0} \int_X \Phi\wedge G_X\nn
\eea

Next we turn our attention to the Kahler potential. We postulate that
\be
\V = \frac{1}{7}\int_X\Phi\wedge\star_7\Phi
\ee
which is the case if $\tilde s^i=0$ \cite{HM} or $s^I=0$ \cite{BW}.
Then the general arguments of \cite{BW} (see also \cite{JG,GP,HM})
show that 
\be
K  = -3\ln\left(\frac{\V}{V_0}\right)
\ee
is the correct Kahler potential. One can also see that
\be
\p_i K  = \frac{i}{2\V}\int_X\omega_i\wedge \star_7\Phi\ ,\qquad
\p_I K  = \frac{i}{2\V}\int_X\varphi_I\wedge \star_7\Phi
\ee
which allows us to evaluate
\bea
g^{i\bar j}\p_i K\bar\p_jK+g^{I\bar J}\p_IK\bar \p_JK &=& 7\nn
g^{i\bar j}\p_iK W\bar \p_j \bar W &=& 2i W\Im W - \frac{i}{2V_0} W\tilde C^I\int_X\varphi_I\wedge d\Phi\nn
g^{I\bar J}\p_IK W\bar\p_J\bar W &=& \frac{1}{2V_0}W\int _X \Phi \wedge d\Phi
+\frac{i}{2V_0}W\tilde C^I\int_X \varphi_I\wedge d\Phi\nn
\eea
Putting this all together gives
\bea
U_{eff} &=&\frac{16\lambda_7^2}{\V}\tilde C^I\tilde C^Jg_{I\bar J}
+\frac{V_0}{16\V^3}g^{I\bar J}\int_X \Phi_I \wedge d\Phi \int_X \Phi_J \wedge d\Phi \nn &&
+\frac{V_0}{4\V^2} \int_X G_X\wedge \star G_X +\frac{4V_0^3}{\V^3}|W|^2\nn
&&+\frac{V_0^2}{\V^3}(\Re W)\int _X \Phi \wedge d\Phi
-\frac{4V_0^3}{\V^3}(\Im W)^2 \nn
&=&\frac{16\lambda_7^2}{\V}\tilde C^I\tilde C^Jg_{IJ}
+\frac{V_0}{4\V^2} \int_X G_X\wedge \star G_X \nn&&
+\frac{V_0}{4\V^3}\left(\frac{1}{2}C^i\int_X \omega_i\wedge G_X  +\frac{1}{2}\tilde C^I\tilde C^J\int_X \varphi_I\wedge d\varphi_J-\tilde P_0\right)^2
\nn
&&
+\frac{V_0}{16\V^3}g^{I\bar J}\int_X \Phi_I \wedge d\Phi \int_X \Phi_J \wedge d\Phi - \frac{V_0}{16\V^3}\left(\int_X \Phi \wedge d\Phi \right)^2
\nn
\eea
The last line can be evaluated more explicitly with the help of  the identity 
$g_{I\bar J} s^I s^J +  g_{i\bar j}\tilde s^i\tilde s^j=\frac{7}{4} $ 
and leads to a potential for
the $\tilde s^i$. In this way we arrive at the final expression for $U_{eff}$
\bea
U_{eff}
&=&\frac{16V_0 \lambda_7^2}{\V}\tilde C^I\tilde C^Jg_{I\bar J}
+\frac{V_0}{4\V^2} \int_X G_X\wedge \star G_X \nn&&
+\frac{V_0}{4\V^3}\left(\frac{1}{2}C^i\int_X \omega_i\wedge G_X  +\frac{1}{2}\tilde C^I\tilde C^J\int_X \varphi_I\wedge d\varphi_J-\tilde P_0\right)^2
\nn
&& 
-\frac{21\lambda_7^2V_0}{\V}- \frac{16\lambda_7^2V_0}{\V}\left(g_{i\bar j}\tilde s^i\tilde s^j 
-(g_{i\bar j}\tilde s^i\tilde s^j )^2\right)
\nn
\eea 
Here we 
find exact agreement with the potential of the previous section if we set
$\tilde s^i=0$. We also see that the $\tilde s^i$  are indeed tachyonic in the 
vacuum corresponding to a weak $G_2$ manifold, \ie\ at $\tilde s^i=0$.
Note that supersymmetry, or more accurately the existence of the 
superpotential,
implies that any tachyonic modes about a supersymmetric solution will
satisfy the Breitenlonher-Freedman bound and do not represent an instability.

Next we can look for supersymetric vacua. These are solutions of 
$D_iW=D_IW=0$;
\bea
0&=&\frac{1}{4V_0}\int_X \omega_i\wedge G_X  +  \frac{iW}{2\V}\int_X\omega_i\wedge \star_7\Phi\nn
0&=& \frac{1}{4V_0}\tilde C^J\int_X\varphi_I\wedge d\varphi_J+\frac{i}{4V_0}\int_X \varphi_I\wedge d\Phi
+  \frac{iW}{2\V}\int_X\varphi_I\wedge \star_7\Phi\nn
\eea
The first equation tells us that either $\Re W=0$ or 
$\int_X\omega_i\wedge \star_7\Phi=0$. 

In the former case we learn from 
the second equation that $d\Phi=0$, \ie\ $s^I=0$ and $\lambda_7=0$. 
This further implies that 
$\tilde C^I=0$. However a little algebra shows that $G_X=0$ and hence
we have a $G_2$ compactification with $\tilde P_0=W=0$.

In the latter  case $\tilde s^i=0$ 
and the first equation gives $G_X=0$.
A little bit of algebra shows that $\tilde C^I = 0$, $\tilde P_0
=-6\lambda_7\V$ and $W  = -2\lambda_7 \V V_0^{-1}$. This is of course just the Freund-Rubin solution (\ref{FR}).

Thus in the presence of fluxes supersymmetry is broken and one is
driven to a global minimum with $M=\tilde C^I=0$ as we mentioned above. 
However
we now see that the metric moduli have a minimum at 
$g_{i\bar j}\tilde s^i\tilde s^j=\frac{1}{2}$ and hence
$g_{I\bar J}s^I s^J=\frac{5}{4}$. In this case the volume and
cosmological constant are fixed to
\be
\V =
\frac{\lambda_7^{-2}}{60}\int_X G_X\wedge \star_7 G_X
\qquad 
U_{eff} = - 600V_0\lambda_7^4\left(\int_X G_X\wedge \star_7 G_X\right)^{-1}
\label{Gvac}
\ee

\section{\sl Modeling Supersymmetry Breaking}

Even if we were able to find the standard model (or a 
supersymmetric/grand-unified   generalization of it) in one of the compactifications discussed above
things would still not be very realistic. In particular these vacua
suffer from a negative vacuum energy density and 
ideally one would like to create vacua with a positive cosmological constant.
Another problem here is that in a typical Freund-Rubin compactification the
Kaluza-Klein scale is of the same order of magnitude as the cosmological
constant. Thus it is of interest to separate these scales so that one can
probe microscopic distances without exciting Kaluza-Klein modes.

In $G_2$ and weak $G_2$ compactifications the phenomenologically relevant
non-Abelian  gauge fields arise on
certain codimension four surfaces $Q\subset X$ and the charged
chiral Fermions arise at co-dimension seven conical singularities
in $X$ that also sit on $Q$ (for a review see \cite{AG}). One method of breaking supersymmetry
in such a scenario would be to imagine that one or more of the
gauge theories located at the singularities dynamically breaks
supersymmetry. This would manifest itself in a positive vacuum
energy located at the singularity. Since we expect that such
breaking effects are due to the Fermions this non-vanishing vacuum
energy should be confined to points in $X$ where there is a conical
singularity.

Therefore to model such effects in the effective action we could add
a term
\be
S_{susy} = -\sum_A  \int d^4x \sqrt{-{}^\star g}\Lambda_A
\ee
where $\Lambda_A$ is a positive vacuum energy and $A=1,...,n$ labels 
the singularity. This vacuum energy 
may well depend
on the moduli of $X$ however since comes from localized sources on $X$
we assume that it is independent of $\V$ and for simplicity we
will also assume that it is independent of the axions.
This results in an extra term appearing in $U_{eff}$
\bea
U_{eff} &=&   -{21  V_0\lambda_7^2\ov \V}
 + \frac{1}{4}{V_0\ov \V^2}\int_X G_X\wedge \star_7 G_X + {\Lambda_{susy}V_0\over \V^2}\nn
&&
+{V_0\over 4 \V^3}\left(\frac{1}{2}C^k \int_X \omega_k\wedge G_X  
+{1\over 2}\int_X C_X\wedge G_X  - P_0\right)^2\nn&&
- \frac{16\lambda_7^2V_0}{\V}\left(g_{i\bar j}\tilde s^i\tilde s^j 
-(g_{i\bar j}\tilde s^i\tilde s^j )^2\right)
\nn
\eea 
where $\Lambda_{susy} =\sum_A\Lambda_A\kappa^{9} $ 
This has a very  similar dependence on $\V$ as the flux
term in $U_{eff}$.

Since there are tachyonic modes about the $\tilde s^i=0$ vacuum, and we are
breaking supersymmetry, we cannot guarantee stability of the $\tilde s^i=0$ vacua (although
imposing the Breitenlohner-Freedman bound would provide perturbative
stability). Therefore 
we will consider vacua where $g_{i\bar j}\tilde s^i\tilde s^j=\frac{1}{2}$.  
There are essentially two cases to consider. In the first case suppose
that there is no flux, \ie\ pure Freund-Rubin. 
One then finds a minimum for $\V $ 
\bea
\V &=& \sqrt{
{7P_0^2\over 300\lambda_7^2}+{7^2\Lambda^2_{susy}\over 25^2\cdot 9^2\lambda_7^4}}+{7\Lambda_{susy}\over 9\cdot 25\lambda_7^2}\nn
U_{eff} &=& - \frac{V_0}{\V^3}\left(\frac{5}{9}\Lambda_{susy}\V 
+ \frac{1}{3}P_0^2\right)\nn
\eea
In the second case we turn on the fluxes (which we have seen breaks 
supersymmetry even if $\Lambda_{susy}=0$). The minimum will 
occur at $M=0$ but now with
\bea
\V &=& 
{14\over 9\cdot 25}\lambda_7^{-2}\Lambda_{susy} + \frac{1}{60}\lambda_7^{-2}\int_X G_X\wedge \star_7 G_X\nn
U_{eff} &=& - \frac{V_0}{\V^2}\left(\frac{5}{9}\Lambda_{susy} 
+ \frac{1}{6}\int_X G_X\wedge \star_7 G_X\right)\nn
\nn
\eea

Thus we see that in this model turning on a positive $\Lambda_{susy}$
raises the volume of the internal space and lowers the magnitude of 
cosmological
constant but it nevertheless  remains negative. One sees that, although 
the supersymmetry breaking contribution to the vacuum energy
density can be made arbitrarily large for 
any given  volume, this is not sufficient to overcome the background 
negative energy density since one  also
finds that the volume is shifted and this in turn  suppresses the contribution 
of $\Lambda_{susy}$ to the  the vacuum energy.
 Note that one cannot make $U_{eff}$ small
by tuning $\Lambda_{susy}$ to a negative value as this will cause the
volume to become negative.

We would like to return to our question about how much we can separate the
Kaluza-Klein and cosmological constant scales. For this we can choose
$V_0$ to be the volume in the vacuum we wish to discuss 
(so that there is no spurious 
conformal factor in the metric ansatz (\ref{gdef})). 
One then sees that in all the vacua with $\Lambda_{susy}=0$,
 $U\sim -\lambda_7^2$. Hence the four-dimensional
cosmological constant is of the same order as $\lambda_7$, which
in turn we expect to be the same order as the Kaluza-Klien scale, 
$\lambda_7\sim \V^{-\frac{1}{7}}$. In order to break this
relationship we need to somehow fine-tune the potential so as to
ensure that the various terms that contribute to $U$ cancel to a
high degree. Unfortunately turning on $\Lambda_{susy}$ does not
seem to enable us to do this.

\section{\sl Comments}

In this paper we discussed the low energy effective action for
M-theory compactified on weak $G_2$ manifolds in the presence of
topologically non-trivial fluxes. This required the introduction of
two types of complex moduli, those associated to massless axions and
those associated to massless metric deformations.
The 
appearance of such fluxes leads to a decay of the Freund-Rubin parameter
mediated by the production of the massless axion modes. 
However there is a non-supersymmetric  global minimum of the 
effective potential in the presence of fluxes which is 
the end point of such a  decay. 
Although it should be born in mind that in a cosmological setting such a  decay
between  spacetimes with negative vacuum energies is likely to end
in  big crunch, rather than pure anti-de Sitter space 
(see \cite{Tom}). It would be interesting to obtain a more explicit 
understanding of this non-supersymmetric solution.

We also  presented the superpotential for the low energy supergravity
and saw that the only supersymmetric solutions correspond to vanishing 
flux; either a pure $G_2$ compactification or a 
pure Freund-Rubin compactification.  Lastly we
discussed a simple model designed to  incorporate the effect of   
supersymmetry breaking by fields localized at codimension seven singularities
in the compact manifold. In this model the cosmological constant can
be increased, but it always remains negative, nor does it enable us
to seperate the Kaluza Klein and cosmological scales. 
Since there is no Bose-Fermi mass degeneracy in a 
supersymmetric anti-de Sitter vacuum  one might hope for phenomenologically 
interesting supersymmetry breaking patterns to arise and it would be
interesting to study in greater detail.

Finally we would also like to mention 
that there are Freund-Rubin solutions with 
topologically trivial fluxes.  These have the form
\be
G_X = \star_7\Phi\ ,\qquad M = -4\lambda_7
\ee
where $\Phi$ is the weak $G_2$ three-form.  
Such solutions were first constructed
long ago \cite{Englert}. Presumably these solutions correspond here
to  setting $G_X=0$ and taking $P_0$ sufficiently negative so that
there is a second extremum at a non-zero value of $\tilde C^I$.
It would be interesting consider a similar analysis by expanding
about such flux backgrounds.

\section*{\sl Acknowledgments}

I would like to thank B. Acharya, F. Denef, M. Douglas, 
R. Minasian, G. Moore and
D. Tong for helpful discussions. This work was supported by  a
PPARC advanced Fellowship as well as the PPARC research 
grant PPA/G/O/2002/00475.


\begin{thebibliography}{10}

\bibitem{Mike}
M.~R.~Douglas,
``The statistics of string / M theory vacua,''
JHEP {\bf 0305}, 046 (2003)
[arXiv:hep-th/0303194].
 
\bibitem{AG}
B.~S.~Acharya and S.~Gukov,
``M theory and Singularities of Exceptional Holonomy Manifolds,''
Phys.\ Rept.\  {\bf 392}, 121 (2004)
[arXiv:hep-th/0409191].


\bibitem{ADV}
B.~S.~Acharya, F.~Denef and R.~Valandro,
``Statistics of M theory vacua,''
arXiv:hep-th/0502060.



\bibitem{ADHL}
B.~S.~Acharya, F.~Denef, C.~Hofman and N.~Lambert,
``Freund-Rubin revisited,''
arXiv:hep-th/0308046.

\bibitem{GVW}
S.~Gukov, C.~Vafa and E.~Witten,
``CFT's from Calabi-Yau four-folds,''
Nucl.\ Phys.\ B {\bf 584}, 69 (2000)
[Erratum-ibid.\ B {\bf 608}, 477 (2001)]
[arXiv:hep-th/9906070].



\bibitem{Sergei}
S.~Gukov,
``Solitons, superpotentials and calibrations,''
Nucl.\ Phys.\ B {\bf 574}, 169 (2000)
[arXiv:hep-th/9911011].



\bibitem{AS}
B.~S.~Acharya and B.~Spence,
``Flux, supersymmetry and M theory on 7-manifolds,''
arXiv:hep-th/0007213.


\bibitem{BW}
C.~Beasley and E.~Witten,
``A note on fluxes and superpotentials in M-theory compactifications on
manifolds of G(2) holonomy,''
JHEP {\bf 0207}, 046 (2002)
[arXiv:hep-th/0203061].

\bibitem{Bobby}
B.~S.~Acharya,
``A moduli fixing mechanism in M theory,''
arXiv:hep-th/0212294.

\bibitem{Minasian}
P.~Kaste, R.~Minasian and A.~Tomasiello,
``Supersymmetric M-theory compactifications with fluxes on  seven-manifolds
and G-structures,''
JHEP {\bf 0307}, 004 (2003)
[arXiv:hep-th/0303127].

\bibitem{Behrndt}
K.~Behrndt and C.~Jeschek,
``Fluxes in M-theory on 7-manifolds and G structures,''
JHEP {\bf 0304}, 002 (2003)
[arXiv:hep-th/0302047];
``Fluxes in M-theory on 7-manifolds: G-structures and superpotential,''
Nucl.\ Phys.\ B {\bf 694}, 99 (2004)
[arXiv:hep-th/0311119].

\bibitem{Lukas}
B.~de Carlos, A.~Lukas and S.~Morris,
``Non-perturbative vacua for M-theory on G(2) manifolds,''
JHEP {\bf 0412}, 018 (2004)
[arXiv:hep-th/0409255].

\bibitem{GL}
T.~W.~Grimm and J.~Louis,
``The effective action of type IIA Calabi-Yau orientifolds,''
arXiv:hep-th/0412277.




\bibitem{HM}
T.~House and A.~Micu,
``M-theory compactifications on manifolds with G(2) structure,''
arXiv:hep-th/0412006.



\bibitem{Greg}
G.~W.~Moore,
``Anomalies, Gauss laws, and page charges in M-theory,''
arXiv:hep-th/0409158.



\bibitem{Ruben} R.~Minasian, private communication.




\bibitem{JG}
J.~A.~Harvey and G.~W.~Moore,
``Superpotentials and membrane instantons,''
arXiv:hep-th/9907026.


\bibitem{GP}
J.~Gutowski and G.~Papadopoulos,
``Moduli spaces and brane solitons for M theory compactifications on  holonomy
G(2) manifolds,''
Nucl.\ Phys.\ B {\bf 615}, 237 (2001)
[arXiv:hep-th/0104105].





\bibitem{Tom}
T.~Banks,
``Heretics of the false vacuum: Gravitational effects on and of vacuum decay.
arXiv:hep-th/0211160.


\bibitem{Englert}
F.~Englert,
``Spontaneous Compactification Of Eleven-Dimensional Supergravity,''
Phys.\ Lett.\ B {\bf 119}, 339 (1982).



\end{thebibliography}
\end{document}